\documentclass[twocolumn,english,aps,superscriptaddress, pra,groupedaddress]{revtex4-1}

\usepackage[latin9]{inputenc}
\usepackage{amsmath}
\usepackage{amssymb}
\usepackage{graphicx}
\usepackage{babel}
\usepackage{mathrsfs}
\usepackage{amsfonts}
\usepackage{epstopdf}
\usepackage{caption}
\usepackage{multirow}

\numberwithin{equation}{section}
\renewcommand\theequation{\arabic{section}.\arabic{equation}}

\begin{document}

\title{Scale Invariant Quantum Dynamics and Universal Quantum Beats \\ in Bose Gases}

\author{Jeff Maki}
\author{Shao-Jian Jiang}
\author{Fei Zhou}

\affiliation{Department of Physics and Astronomy, University of British Columbia, Vancouver V6T 1Z1, Canada}

\date{Mar. 6, 2014}

\begin{abstract}

 We study the signature of scale invariance in the far-from-equilibrium quantum dynamics of two dimensional Bose gases. We show that the density profile displays a scale invariant logarithmic singularity near the center. In addition, the density oscillates due to quantum beats with universal structures. Namely, the frequencies of the beats can be connected with one another by a universal discrete scale transformation induced by the scale invariance. The experimental applicability of these results is then discussed.

\end{abstract}

\maketitle

\section{Introduction}
One of the most important tools in physics research is to explore the consequences and implications of symmetries in a physical system, i.e. the invariance of a system under a family of transformations. Any symmetry or invariance can lead to a significant reduction of the complexity of a problem, possibly rendering it solvable. Therefore it is crucial to investigate and understand the features that are invariant due to symmetry transformations. One example of great interest is the symmetry associated with scale invariance. This symmetry has been essential in the study of complicated systems such as the thermodynamics \cite{Wilson83, Sachdev} and dynamics \cite{Hohenburg_Rev} near critical points, and biological complexes \cite{Brown00}. Without taking into account this symmetry, these problems would be otherwise intractable. Although scale invariance has had resounding success in studying the energetics and critical dynamics of these rather difficult systems, the role of scale invariance on far-from-equilibrium coherent quantum dynamics has yet to be  fully understood. In this work we are specifically interested in the scale invariant far-from-equilibrium quantum dynamics, where the governing equations of a system remain unchanged when subject to dilations of the spatial and temporal coordinates:

\begin{align}
 \vec{x}_i'  &= b  \vec{x}_i   & i &=1,2,...,N   &  t' &= b^2 t.
\label{scaling}     
\end{align}

\noindent Below we use $\left\lbrace \vec{x}_i \right\rbrace$ to denote the coordinates of the $N$-particles with $i = 1,2,...,N$, and $b$ as the scaling factor.

One  very promising platform to study the role of scale invariance in far-from-equilibrium dynamics is cold atom systems.  Cold atom systems are unique in their tunability present in experiments. It is possible to prepare cold atom systems with a wide range of initial conditions and interaction parameters. This control has led to both experimental and theoretical studies concerning  a wide range of dynamical phenomena, such as the dynamics of expansion and collapse \cite{Bosenova, Castin_Dum,Kagan96, Stoof97}, breathing modes \cite{Mewes96, Stringari96}, solitons \cite{Sengstock,Hulet, Reatto02}, and quench dynamics \cite{Bosenova, Makotyn,Radzihovsky16}. A few theoretical efforts have also been made to understand far-from-equilibrium coherent quantum dynamics in Fermion superfluids \cite{Spivak04, Gurarie}.

One of the simplest cold atom systems to exhibit scale invariance is the two dimensional Bose gas with contact interactions of strength $-g$ \cite{csi_note}. Under the spatial dilation given by Eq.~(\ref{scaling}), the many body Hamiltonian for this system:

\begin{equation}
H = \sum_i \frac{\vec{P}_i^2}{2} - \frac{g}{2}\sum_{i \neq j} \delta^{(2)}(\vec{x}_i - \vec{x}_j),
\label{hamiltonian}
\end{equation}

\noindent transforms as $H' = b^{-2} H$, where $\hbar$ and the atomic mass have been set to unity; $m_0 = 1$. If one simultaneously considers the temporal dilation of Eq.~(\ref{scaling}), one can verify that the Schrodinger equation for the $N$ particle  many body wave function, $\Psi(\left \lbrace \vec{r}_i \right\rbrace, \lbrace \lambda \rbrace, t)$:

\begin{equation}
i \partial_t \Psi(\left\lbrace \vec{x}_i \right\rbrace, \lbrace \lambda \rbrace, t) = H \Psi(\left\lbrace \vec{x}_i \right\rbrace, \lbrace \lambda \rbrace, t)
\label{Schrodinger}
\end{equation}

 \noindent is invariant. From these general observations it is appealing to argue that in two dimensions the interacting Bose gas is scale invariant.
 
The additional parameters, $\lbrace \lambda \rbrace$, are the length scales set by the initial conditions. Although the Schrodinger equation is scale invariant, it is still necessary to impose initial conditions on the dynamics. These initial length scales break the scale invariance of the problem. In order to restore scale  invariance, it is necessary to rescale the initial conditions alongside the spatial and temporal coordinates, Eq.~(\ref{scaling}). As a result, the scale invariance relates the dynamics not only at different points in space and time, but to different initial conditions:
 
 \begin{equation}
 \Psi( b \left\lbrace \vec{x}_i \right\rbrace, b \lbrace \lambda \rbrace, b^2t ) = b^{-N} \Psi( \left\lbrace \vec{x}_i \right\rbrace, \lbrace \lambda \rbrace, t),
 \label{eq:psi_homogeneous}
 \end{equation}

\noindent which means that the many body wave function is a homogeneous function. The scaling exponent of the many body wave function is fixed  to be $-N$ in order for the normalization to be independent of scale. The fact that the many body wave function is a homogeneous function, implies that the unitary evolution of a quantum state can be effectively described by a simple scale transformation, i.e. Eq.~(\ref{scaling}). By extension it follows that all physical observables will be homogeneous functions. The specific form of the homogeneous function and the value of the scaling exponent for a given observable is the subject of interest below.

In this article we study the signatures of scale invariance on the quantum dynamics of a two dimensional Bose Einstein condensate with attractive contact interactions. Although the scale invariance has been studied in the thermodynamics of this system \cite{Dalibard07, Chin11}, the role of scale invariance in the far-from-equilibrium dynamics has yet to be determined. We restrict ourselves to an isotropic single parameter scaling solution and show how the scale invariance leads to distinct features present in the density profile of the gas. First, at short distances, the spatial profile is dictated by the presence of a logarithmic singularity in the density. Secondly, this density profile will undergo oscillations, the frequencies of which satisfy a robust discrete scaling relation. Both these effects are universal in the sense that they  do not depend on the initial conditions of the condensate and are valid for a wide range of interaction strengths. We then conclude with a discussion on the experimental implications of this work.

\section{Dynamics and the Quantum Variational Method}

Consider the expectation value of the density operator $\hat{\rho}(\vec{r}) = \hat{\phi}^{\dagger}(\vec{r}) \hat{\phi}(\vec{r})$ where $\hat{\phi}^{(\dagger)}(\vec{r})$ is the second quantized annihilation (creation) operator. The unitary evolution of the density is given by:

\begin{widetext}
\begin{eqnarray}
\rho(\vec{r},t) &=&  \left\langle \psi_0 |  e^{i H t}\hat{\rho}(\vec{r}) e^{-i H t}  | \psi_0 \right\rangle = \nonumber \\
& & \frac{\int D \phi (\vec{x})D \phi '(\vec{x}) \left\langle \psi_0  |e^{i H t}| \lbrace \phi(\vec{x} ) \rbrace \right\rangle  \left\langle \lbrace \phi ' (\vec{x}) \rbrace |e^{-i H t}| \psi_0 \right\rangle \phi ^*(\vec{r}) \phi '(\vec{r}) \langle \lbrace \phi(\vec{x}) \rbrace | \lbrace \phi'(\vec{x}) \rbrace \rangle}{\int D\phi (\vec{x}) | \left\langle \lbrace \phi (\vec{r}) \rbrace | e^{-i H t} | \psi_0 \right\rangle |^2}, 
\label{eq:TimeEvolved}
\end{eqnarray}
\end{widetext}


\noindent  where $|\psi_0 \rangle$ is the initial state of the system, $H$ is the second quantized form of Eq.~(\ref{hamiltonian}) and $| \lbrace \phi(\vec{x}) \rbrace \rangle $ is a many body coherent state defined as the eigenstate of the annihilation operator $\hat{\phi}(\vec{r})$: $\hat{\phi}(\vec{r}) | \lbrace \phi(\vec{x}) \rbrace \rangle = \phi(\vec{r}) | \lbrace \phi(\vec{x}) \rbrace \rangle$ \cite{Negele_Orland}. In general one can write the transition amplitude, $ \left\langle \psi_0  |e^{-i H t}| \lbrace \phi(\vec{x}) \rbrace \right\rangle$, in terms of a functional integral \cite{Feynman} over complex scalar fields. In practice this offers no simplification. For this reason the theory of dynamics is often restricted to semiclassical approaches, numerical methods \cite{Castin_Dum, Kagan96, Stoof97}, or the Heisenberg equations of motion \cite{Deng}. However, one can make two further simplifications for low dimensions and dense condensates \cite{Maki14}. The first is to note that for dense condensates $\langle \lbrace \phi(\vec{x}) \rbrace | \lbrace \phi'(\vec{x}) \rbrace \rangle = \delta \left( \lbrace \phi'(\vec{x}) \rbrace - \lbrace \phi(\vec{x}) \rbrace \right)$, which requires that the two field configurations are identical (see Appendix \ref{sec:appendix_A} for more details). This is possible since small deviations in the many body coherent states lead to nearly orthogonal states. Secondly, one can express $\phi(\vec{x})$ as $\phi(\vec{x}) = \phi_{0}(\vec{x}) + \delta \phi(\vec{x})$, where $\delta \phi(\vec{x})$ represents the anisotropic many body fluctuations which arise from short wave length - fast phonons, and $\phi_{0}(\vec{x})$ the isotropic long wave length - slow degrees of freedom. The isotropic slow degrees of freedom can be described using a single parameter ansatsz:

\begin{equation}
 |\phi_{0}(\vec{x})|^2 = \rho_{\lambda}(\vec{x}) = \frac{N}{ \lambda^2} f\left( \frac{x}{\lambda} \right).
\label{eq:ansatz}
\end{equation}

\noindent The phase of $\phi_{0}(\vec{x})$, is chosen to satisfy the conservation law \cite{Maki14} and is of little importance for the rest of the discussion. The quantity $\lambda$ parametrizes the slowly evolving field and $f(x)$ is a smooth normalizable function that is regular at the origin. This ansatsz is motivated by the generic scale invariance of the many body Schrodinger equation and the initial conditions of an inhomogeneous condensate. For more details we refer the reader to Appendix \ref{sec:Appendix_B}.

The separation of the slow isotropic and fast anisotropic degrees of freedom leads to a controllable expansion of the many body fluctuations valid in the limit of dense condensates. After substituting the split form of $\phi(\vec{x})$ into Eq.~(\ref{eq:TimeEvolved}) and integrating out $\delta \phi(\vec{x})$  (see Appendix \ref{sec:Appendix_C}), Eq.~(\ref{eq:TimeEvolved}) reduces to:

\begin{equation}
\rho(\vec{r},t) = \langle \rho_{\lambda}(\vec{r}) \rangle (t) = \frac{\int d\lambda \ \rho_{\lambda}(\vec{r})|\psi(\lambda,t)|^2 }{\int d \lambda \ |\psi(\lambda,t)|^2}.
\label{eq:final_avg}
\end{equation}

\noindent  The last equality in Eq.~(\ref{eq:final_avg}) indicates that the density at any given time, $t$, can be obtained by simply averaging $\rho_{\lambda}(\vec{r})$ in Eq.~(\ref{eq:ansatz}) over $|\psi(\lambda,t)|^2$, or $\langle \rho_{\lambda}(\vec{r}) \rangle (t)$.

The quantity $\psi(\lambda,t)$ is the transition amplitude for the slow degrees of freedom, $\langle \lambda | e^{-i H_{\lambda} t} | \psi_0\rangle$. The Hamiltonian governing this transition amplitude  is given by:

\begin{equation}
H_{\lambda} = \frac{\hat{P}_{\lambda}^2}{2m}  - \frac{V}{2 \hat{\lambda}^2} + \delta H_{\lambda},
\label{eq:effective_hamiltonian}
\end{equation}

\noindent where $\delta H_{\lambda}$ is a small correction due to the anisotropic many body fluctuations \cite{Maki14}, the effect of which will be discussed towards the end. In this effective description, $| \lambda \rangle$ is an eigenstate of the operator $\hat{\lambda}$: $\hat{\lambda} | \lambda \rangle = \lambda | \lambda \rangle$, representing a condensate with wave function $\phi_{0}(\vec{x})$, while $\hat{P}_{\lambda}$ is the momentum conjugate to $\hat{\lambda}$. The constants $m$ and $V$ are given by $C_1 N$ and $ C_3 g N^2- C_2 N$ respectively, where $C_1$, $C_2$, and $C_3$ have been calculated previously \cite{Maki14}. Finally, it is important to note that Eq.~(\ref{eq:effective_hamiltonian}) with $\delta H_{\lambda} =0$ is in fact scale invariant, reflecting the symmetry in the original microscopic Hamiltonian, Eq.~(\ref{hamiltonian}).

The spectrum of Eq.~(\ref{eq:effective_hamiltonian}) with $\delta H_{\lambda} = 0$ consists of a continuous set of scattering states, $\psi_{s}^{(1)}$ and $\psi_{s}^{(2)}$, with energies $E = \frac{k^2}{2m}$, and a discrete set of bound states, $\psi_b$, with energies $E_n = - \frac{k_n^2}{2m}$, where (up to normalization factors):

\begin{align}
\psi_{s}^{(1)} &= \text{Re} \ \sqrt{k \lambda} J_{a}(k \lambda), & \psi_{s}^{(2)} &= \text{Re} \ \sqrt{k \lambda} Y_{a}(k \lambda), \nonumber \\
\psi_b &= \sqrt{k \lambda} K_{a}(k_n \lambda), & k_n &= k_0 \exp\left(\frac{- n \pi}{\sqrt{mV}}\right).
\label{eq:states}
\end{align}

\noindent The functions $J_{a}(x)$, $Y_{a}(x)$, and $K_{a}(x)$ are the Bessel J, Bessel Y and modified Bessel K functions of order $a = i \sqrt{m V-1/4}$, respectively, and  $n = 1,2,3...$ Here we focus on condensates with $mV > 1/4$ and $g \ll 1$, or from the discussion after Eq.~(\ref{eq:effective_hamiltonian}), $C_2 / C_3 N < g  \ll 1$. The length scale $k_0$ specifically depends on the Ultra-violet (UV) features of the problem which regularize the singular $\lambda^{-2}$ potential. The presence of this UV length scale effectively converts the continuous scale invariance to a discrete scale invariance. This results in a bound state spectrum which is equally spaced on a logarithmic scale. It is important to note that both the scattering and bound state eigenfunctions have an envelope that depletes as $(k \lambda)^{1/2}$ as $\lambda \rightarrow 0$. This feature is robust and depends only on the scale invariance of Eq.~(\ref{eq:effective_hamiltonian}).


\section{Quantum Dynamics of the Condensate}

At this stage one can consider the dynamics of a condensate which is initially prepared with size $\lambda_0$. The initial amplitude can be represented as:

\begin{equation}
\psi(\lambda, t=0) = \langle \lambda | \psi_0 \rangle = \frac{1}{\left( \pi \right)^{1/4} \sqrt{\sigma}} e^{-\frac{(\lambda-\lambda_0)^2}{2 \sigma^2}},
\label{eq:initial_conditions}
\end{equation} 

\noindent where the spreading, $\sigma$, is fixed by requiring that the energy of the effective model is identical to the microscopic model: $\sigma = \lambda_0 / (\sqrt{C_1} N)$. Below we present our numerical solutions of Eqs.~(\ref{eq:final_avg}) and (\ref{eq:effective_hamiltonian}) with the initial state  given in  Eq.~(\ref{eq:initial_conditions}).

\begin{figure}
\includegraphics[scale=0.45, trim={0.5cm 0 0 0},clip]{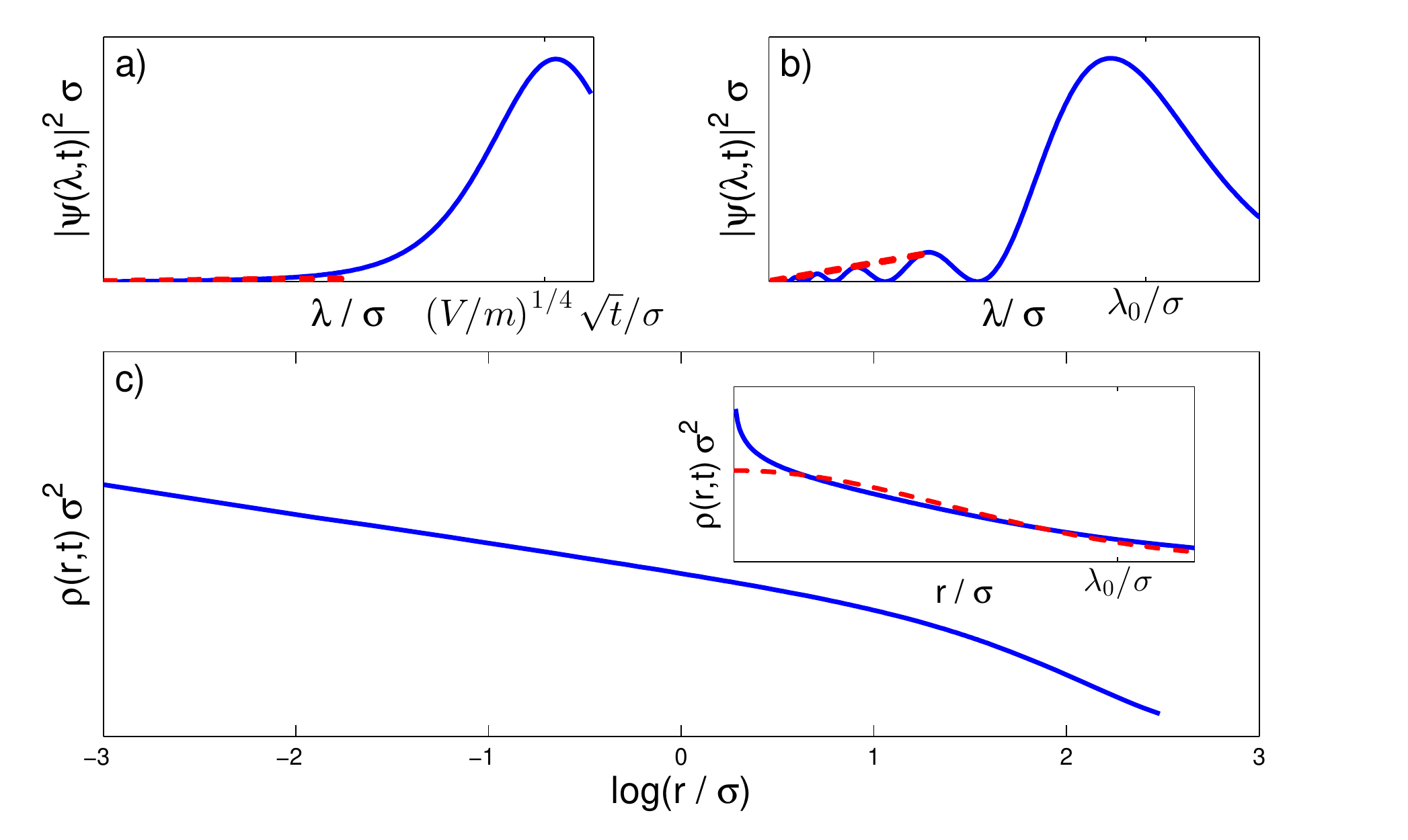}
\caption{The numerical solution of the probability density, $|\psi(\lambda,t)|^2$, and the resulting density profile, Eq. (\ref{eq:final_avg}). a) For $ \lambda \ll \lambda_{sc}^{(1)}(t)$ (only the scattering state contribution is shown, see main text) when $\lambda_0/ \sigma = 50$, $mV$ $= 50$ and $t / (m \sigma^2) =1000$. b) For  $\lambda \ll \lambda_0$ when $\lambda_0 / \sigma = 10$, $m V=$ $ 27.2$ and $t / (m \sigma^2)=1000$. The linear depletion in the probability density is specifically shown by the red dashed lines.  c) The density profile as $r \rightarrow 0$, (Eq.(\ref{eq:rho_r_scaling}),  blue solid) and the semi-classical solution (see main text, red dashed).}
\label{fig:attractive}
\end{figure}
In order to evaluate the transition amplitude at time $t$ it is necessary to examine how the initial state in Eq.~(\ref{eq:initial_conditions}) is projected into the complete set of eigenstates of $H_{\lambda}$ given in Eq.~(\ref{eq:states}). The amount of probability projected into the bound states depends on the ratio of the potential energy to the kinetic energy, $\sqrt{mV}/N$, or from the discussion after Eq.~(\ref{eq:effective_hamiltonian}), $ \sqrt{C_3 g N-C_2}$. The more kinetic (potential) energy the system possesses, the more probability will be concentrated in the scattering (bound) states. Once the projection of the initial state is known, the unitary evolution can be carried out to obtain the probability density, $|\psi(\lambda,t)|^2$. We focus on the limit when the time $t \gg \sqrt{m/V} \lambda_0^2$ and present the probability density in Figs.~(\ref{fig:attractive} a-b).


It is now possible to examine the main quantum effect by considering the average of $\hat{\lambda}^{-2}$ over $|\psi(\lambda,t)|^2$, $\langle \hat{\lambda}^{-2} \rangle(t)$. First one might consider approximating $ \langle \hat{\lambda}^{-2}(t) \rangle$ with $(\lambda_{sc}^{(1)}(t))^{-2}$, where $\lambda_{sc}^{(1)}(t)$ is the most probable value of $\lambda$ in $|\psi(\lambda,t)|^2$ \cite{most_probable_note}; which length scale also represents the semiclassical solution of $H_{\lambda}$ with zero energy: $\lambda_{sc}^{(1)}(t) \sim (V/m)^{1/4} \sqrt{t}$. One then finds that $\rho(0,t) = \rho_{\lambda}(0)$, where $\rho_{\lambda}(\vec{r})$ is given in Eq.~(\ref{eq:ansatz}) and $\lambda = \lambda_{sc}^{(1)}(t)$. The semiclassical solution to the density is finite at the origin since $f(x)$ in Eq.~(\ref{eq:ansatz}) is regular at $x=0$, and is the result one would obtain by employing the hydrodynamical methods \cite{Castin_Dum,Kagan96, Stoof97}. However, our calculations show  that $\langle \hat{\lambda}^{-2} \rangle$ is actually dominated by contributions at small $\lambda$ far away from the most probable value in $|\psi(\lambda,t)|^2$ \cite{sc_vs_quantum_note}.

These anomalous contributions from small $\lambda$ alter the density profile at length scales $\lambda_0 \ll r \ll \lambda_{sc}^{(1)}(t)$. The behaviour of the density at these length scales will be governed by the depletion of the scattering eigenstates, which is explicitly shown in Fig.~(\ref{fig:attractive} a). In this limit $|\psi(\lambda,t)|^2$ depletes linearly and following Eq.~(\ref{eq:final_avg}), it results in a logarithmic singularity in the density profile:

\begin{equation}
\lim_{r / \sqrt{t} \to 0} \rho(\vec{r},t)  = \frac{1}{\pi}  \frac{m}{V} \frac{\lambda_0^2}{t^2}\log^{\alpha} \left( \frac{\sqrt{t}}{r}\right),
\label{eq:rho_r_scaling}
\end{equation}

\noindent as shown in Fig.~(\ref{fig:attractive} c). The power of the logarithm $\alpha = 1$ for the  scale invariant system under consideration. Note that Eq.~(\ref{eq:rho_r_scaling}) obeys: $\rho(b \vec{r}, b^2 t) \sim b^{-2 + \eta} \rho( \vec{r},t)$, fully consistent with the general discussion in the introduction. The exponent $\eta$ is known as the anomalous dimension, and in this  system $\eta = -2$.

\begin{figure}
\includegraphics[scale = 0.46, trim={0.2cm 0 0 0},clip]{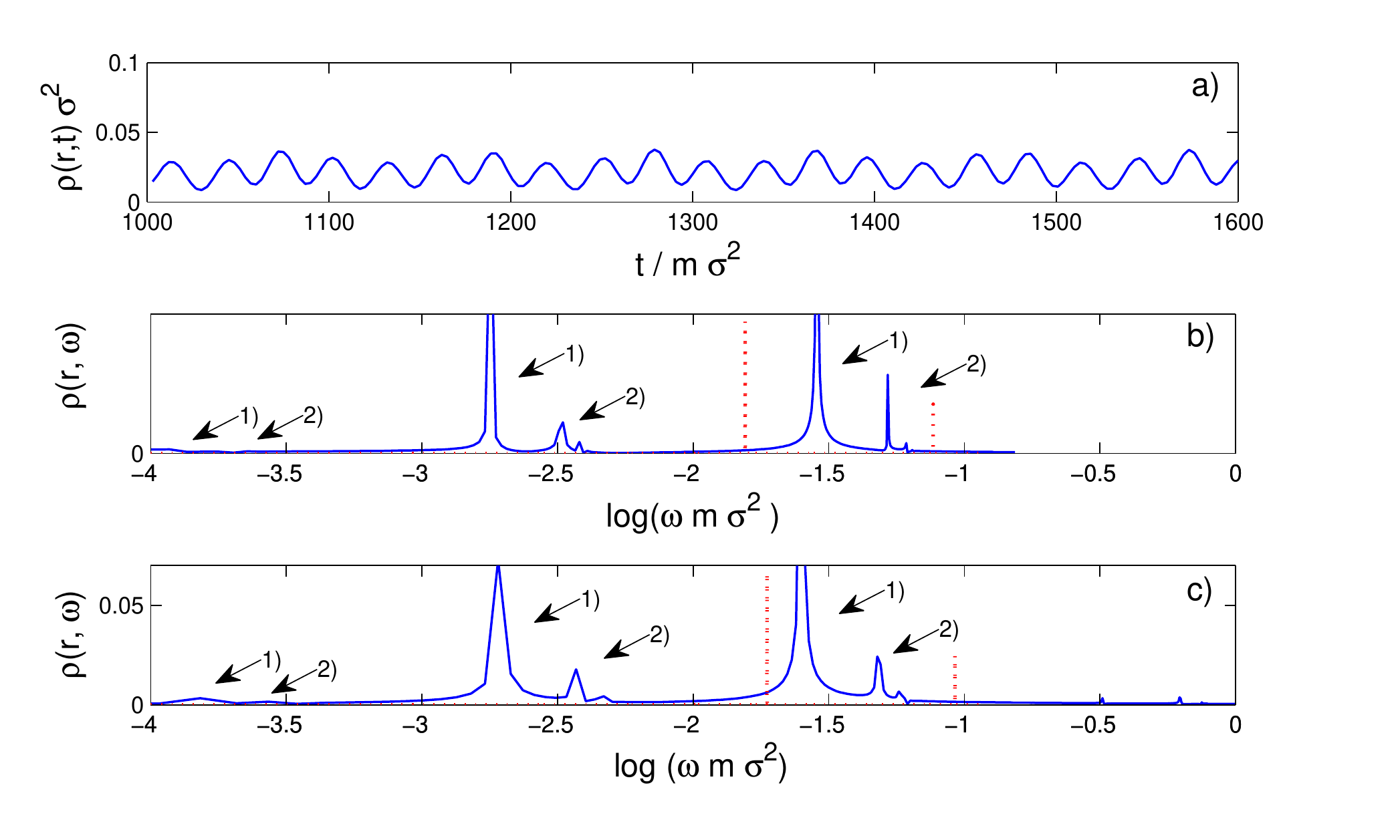}
\caption{a) The temporal evolution of the density profile at a fixed position $r \ll \lambda_0$. For this calculation $r / \sigma = 0.1$, $m V= $ $27.2$, $\lambda_0 / \sigma = 10$. b) The frequency spectrum  (see Eq. (\ref{eq:freq_scaling})), blue solid line) is shown alongside the semi-classical solution (see main text, red dashed line). Only two families are shown explicitly with labels $1)$ and $2)$ corresponding to families with $\nu = 1$ and $\nu=2$, respectively. c) The spectra for $r / \sigma = 0.1$, $m V = 32$ and $\lambda_0 / \sigma = 10$.}
\label{fig:temporal}
\end{figure}

At length scales $\lambda \ll \lambda_0$, the dynamics will be governed by the bound state contribution to the transition amplitude as shown in Fig.~(\ref{fig:attractive} b). There will still be a logarithmic singularity now regulated by $\lambda_0$ due to the depletion of the transition amplitude.  However, the pre-factor of $\lambda_0^2 / t^2$ will be replaced by an oscillatory function. These oscillations are  due to the interference, or beating, with different bound states and is shown in Fig.~(\ref{fig:temporal} a). In Figs.~(\ref{fig:temporal} b-c),  the frequency spectrum of these quantum beats is shown for two interaction strengths. The frequencies of the quantum beats are: $\omega_{n,\nu} = E_{n+\nu} - E_{n}$, with $E_{n}$ given by Eq.~(\ref{eq:states}), and $n, \nu = 1,2,3...$ The exact location of these beat frequencies and their spectral weight will specifically depend on the UV parameter and initial conditions, $\lambda_0$. However, the effect of the induced discrete scale invariance of the system is manifest in the organization of these frequencies. From Eq.~(\ref{eq:states}), the beat frequencies are:
 
\begin{equation}
\log \left( \omega_{n,\nu} m \sigma^2 \right) = \log\left(\frac{k_0^2 \sigma^2}{2}\right) - \frac{2\pi}{\sqrt{mV}}n - \log \left(1 - e^{\frac{2\pi \nu}{\sqrt{mV}}}\right).
\label{eq:freq_scaling}
\end{equation}

\noindent Eq.~(\ref{eq:freq_scaling}) indicates that one can organize the frequency spectrum into a series of families specified by the fixed parameter $\nu$. In this logarithmic scale, the frequencies in each family with given $\nu$ will be equally spaced from one another by an amount: $\frac{2\pi}{\sqrt{mV}}$. The spacing between family members of different $n$ is a universal quantity and is independent of the UV parameter and initial conditions $\lambda_0$. In practice, there will be many families present, each shifted with respect to one another but with the same intra-family spacing. The overall shift between adjacent families, $\nu$ and $\nu +1$, is also universal, as seen by the third term in Eq.~(\ref{eq:freq_scaling}). Our numerical solutions shown in Figs.~(\ref{fig:temporal} b-c) are completely consistent with this general analysis.
 
The universal scaling found in Eq.~(\ref{eq:freq_scaling}) is to be contrasted with the semi-classical solution for a Bose gas with an initial size $\lambda_0$. The classical motion of the size $\lambda_{sc}(t)$ satisfies the equation of motion:

\begin{equation}
m \ddot{\lambda}_{sc}(t) = -\frac{V}{\lambda_{sc}^3(t)},
\label{eq:classical_eom}
\end{equation}

\noindent which is simply the classical motion for a particle subject to an attractive $\lambda^{-2}$ potential. For a condensate with an initial size $\lambda_0$, the solution to Eq.~(\ref{eq:classical_eom}) is:

\begin{eqnarray}
\rho(\vec{r},t) &=& \rho_{\lambda_{sc}}(\vec{r}) =  \frac{N}{\lambda^2_{sc}(t)} f\left( \frac{r}{\lambda_{sc}(t)}\right), \nonumber \\
\lambda_{sc}(t) &=& = \lambda_0  \sqrt{1 - (2(t-nT)/T)^2},
\end{eqnarray}

\noindent for $t \in \left[nT -T/2 ,nT + T/2 \right]$ with $n = 0,1,2,...$ and period, $T$:

\begin{equation}
T= 2\lambda_0^2 \sqrt{\frac{m}{V}}.
\end{equation} 

\noindent We note that the period depends only on the initial conditions of the problem, which is a consequence of the scale invariance of the system. Since this solution oscillates with a period $T$, the frequency spectrum only contains frequencies $\omega_n = 2 \pi n / T$ and $n = 1,2,3,...$ These points are shown alongside the quantum frequency spectrum in Figs.~(\ref{fig:temporal} b-c). 

\begin{figure}
\includegraphics[scale=0.45]{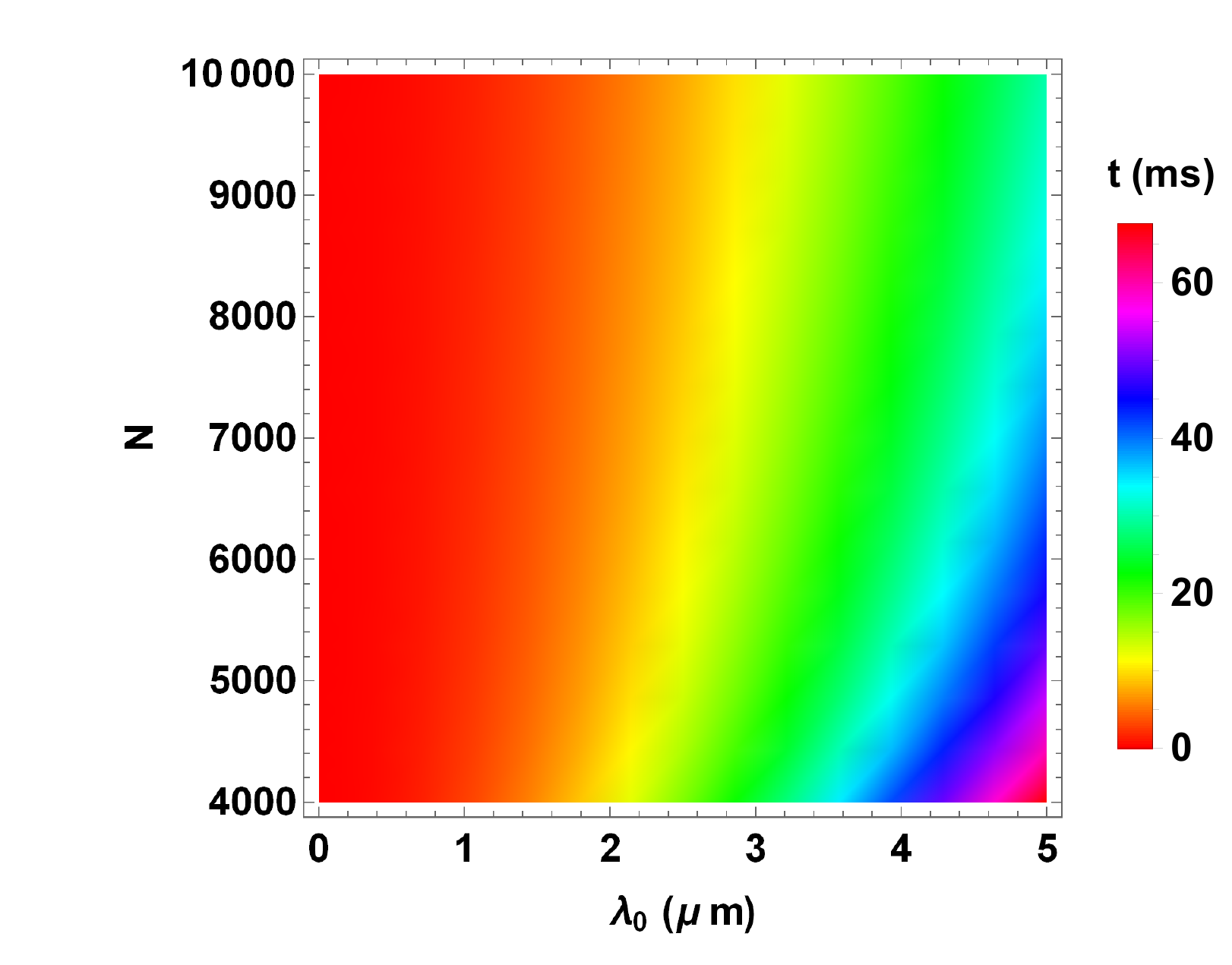}
\caption{The period of a quantum beat. Here we chose $\omega_{n,\nu}$ to be the most dominating peak in Fig.~(\ref{fig:temporal}) for a variety of $N$ and $\lambda_0$. This plot assumes a condensate of $^{133} Cs$ atoms with interaction $g \approx 0.01$.}
\label{fig:temporal_tq}
\end{figure}

\section{Effects of Fluctuations}

The results found in Eqs.~(\ref{eq:rho_r_scaling}) and (\ref{eq:freq_scaling}) neglect higher order fluctuations, that is when $\delta H_{\lambda}$ is set to zero. These fluctuations correspond to corrections which are of higher order in $g$ and can be expanded perturbatively in the limit $C_2 / C_3 N < g \ll 1$. These corrections have two main effects which we will now discuss.

The first effect of the fluctuations is to generate an imaginary correction to the Hamiltonian. The fact that our Hamiltonian has an imaginary contribution is easily understood as the result of the coupling between the long wave length isotropic degrees of freedom and anisotropic fluctuations - phonons. This effect introduces a term $i \ \text{Im} \ \delta H$ to $H_{\lambda}$ where:

\begin{equation}
\text{Im} \ \delta H = \frac{g^2 C_4 N^2}{2 \lambda^2}.
\end{equation}

\noindent This correction is suppressed by an additional factor of $g$. All the eigenstates of energy $E_n$ now acquire a finite lifetime of order $\frac{1}{g E_n}$. This implies that the beat frequencies in the spectrum, Eq.~(\ref{eq:freq_scaling}), will have non-zero widths. The widths associated with the beats with large $\nu$,  $\nu \gg \sqrt{m V}/(2 \pi),$ will be of $O\left(g\right)$ and can be neglected.

The second effect of the fluctuations is to renormalize the parameters $C_1$, $C_2$, and $C_3$, and the coupling constant $g$.  The renormalization of the coefficients $C_1$, $C_2$ and $C_3$, do not qualitatively alter the physics present in the semiclassical model. However, the fluctuations have a more dramatic effect on $g$; the fluctuations replace the bare interaction strength with a function that depends on the UV scale, $k_0$ \cite{Popov, Petrov00, Hammer04, Zhou13, fluct_ref}. This effect explicitly breaks the scale invariance of the system and is known as a quantum anomaly. The effects of the renormalization are are of order $O\left(g \log (k_0 \lambda)\right)$, which is small in the limit under consideration. As a result the scale invariant Hamiltonian found in Eq.~(\ref{eq:effective_hamiltonian}) is a good approximation for the dynamics.

\section{Discussion}
Practically, to observe the log singularity and the universal quantum beats discussed in Figs.~(\ref{fig:attractive}) and (\ref{fig:temporal}), it is 1most convenient to work in the limit when $Ng$ is not too large.  As an example, we consider an experimental set up similar to Ref.~\cite{Chin11}, where $^{133}Cs$ atoms were placed in a two dimensional trap with $g \approx 0.01$. The beats typically occur for $t > \sqrt{V/m} \lambda_0^2$ (see discussion after Eq.~(\ref{eq:freq_scaling})). A contour plot showing the time scale for a single quantum beat for various $N$ and $\lambda_0$ is provided in Fig.~(\ref{fig:temporal_tq}).

In this article we have discussed the role of scale invariance in the quantum dynamics of two dimensional Bose gases. The manifestations of scale invariance in the far-from-equilibrium dynamics, Eqs.(\ref{eq:rho_r_scaling}) and (\ref{eq:freq_scaling}), are accessible to current cold atom experiments. The quantum variational method developed here allows one to perform a controllable calculation of the dynamics in the limit of dense condensates in low dimensions.  This method is quite general, and can be applied to a wide number of systems. In the future we plan to extend this work to study the dynamics of Fermi gases in low dimensions.

This work was supported by NSERC (Canada) and the Canadian Institute for Advanced Research. The authors would like to thank Ian Affleck, Frederic Chevy, and David Feder for helpful discussions.

\appendix

\numberwithin{equation}{section}
\renewcommand\theequation{\Alph{section}.\arabic{equation}}

\section{Quantum Dynamics of the Density Operator}
\label{sec:appendix_A}

In the following appendices, we provide a detailed derivation of Eq.~(\ref{eq:final_avg}) starting from Eq.~(\ref{eq:TimeEvolved}):

\begin{widetext}
\begin{eqnarray}
\rho(\vec{r},t) &=&  \left\langle \psi_0 |  e^{i H t}\hat{\rho}(\vec{r}) e^{-i H t}  | \psi_0 \right\rangle = \nonumber \\
& & \frac{\int D \phi (\vec{x})D \phi '(\vec{x}) \left\langle \psi_0  |e^{i H t}| \lbrace \phi(\vec{x} ) \rbrace \right\rangle  \left\langle \lbrace \phi ' (\vec{x}) \rbrace |e^{-i H t}| \psi_0 \right\rangle \phi ^*(\vec{r}) \phi '(\vec{r}) \langle \lbrace \phi(\vec{x}) \rbrace | \lbrace \phi'(\vec{x}) \rbrace \rangle}{\int D\phi (\vec{x}) | \left\langle \lbrace \phi (\vec{r}) \rbrace | e^{-i H t} | \psi_0 \right\rangle |^2}. \nonumber
\label{eq:appendix_TimeEvolved}
\end{eqnarray}
\end{widetext}

\noindent The states $| \lbrace \phi(\vec{x} \rbrace \rangle$ are the eigenstates of the annihilation operator $\hat{\phi}(\vec{r})$: $\hat{\phi}(\vec{r}) | \lbrace \phi(\vec{x} \rbrace \rangle = \phi(\vec{r}) | \lbrace \phi(\vec{x} \rbrace \rangle$, defined on a discretized lattice with sites: $\vec{x}$. For the current system, we are only interested in the coherent states normalized to the number of particles $N$. That is, in the continuum limit: $\int d^2 x |\phi(\vec{x})|^2 = N$.

The matrix element $\left\langle \lbrace \phi(\vec{x}) \rbrace |e^{-i H t}| \psi_0 \right\rangle$ can be written in terms of a functional integral \cite{Feynman}:

\begin{equation}
\left\langle \lbrace \phi (\vec{x}) \rbrace |e^{-i H t}| \psi_0 \right\rangle = \int' D \phi e^{i S},
\label{eq:appendix_matrix_element}
\end{equation}

\noindent where $S$ is the action for a non-relativistic Bose gas:

\begin{eqnarray}
S &=& \int d^2x \int_0^t dt' \phi^*(\vec{x},t) i \partial_t \phi(\vec{x},t) + \phi^*(\vec{x},t) \frac{\nabla^2}{2} \phi(\vec{x},t) \nonumber \\
& & - \frac{g}{2} |\phi(\vec{x},t)|^4
\label{eq:appendix_action}
\end{eqnarray}

\noindent and  $\int' D \phi$ denotes the sum over all field configurations $\phi(\vec{x},t)$ which satisfy the following boundary conditions:

\begin{eqnarray}
\phi(\vec{x},T) &=& \phi(\vec{x}), \nonumber \\
\phi(\vec{x},0) &=& \psi_0(\vec{x}). 
\end{eqnarray}

When the number of particles at each point in our discretized space is large $|\phi(\vec{x})|^2 \gg 1$, it is possible to simplify Eq.~(\ref{eq:TimeEvolved}) by noting that the overlap between two coherent states approaches a functional delta function:

\begin{eqnarray}
\langle \lbrace \phi(\vec{x}) \rbrace | \lbrace \phi'(\vec{x}) \rbrace \rangle &=& e^{ \int d^2 x \left(\phi^*(\vec{x})\phi'(\vec{x})- \frac{1}{2}|\phi(\vec{x})|^2 - \frac{1}{2}|\phi'(\vec{x})|^2 \right)} \nonumber \\
& \approx & \Pi_{\vec{x}} \ \delta \left( \phi(\vec{x}) - \phi'(\vec{x}) \right) \nonumber \\
&\equiv& \delta \left( \lbrace \phi'(\vec{x}) \rbrace - \lbrace \phi(\vec{x}) \rbrace \right).
\label{eq:appendix_overlap}
\end{eqnarray}

\noindent Eq.~(\ref{eq:appendix_overlap}) states that the value of the two fields $\phi(\vec{x})$ and $\phi'(\vec{x})$ are equivalent at each point in space. In the continuum limit, this is equivalent to a delta function enforcing the two field configurations to be identical. This result simplifies Eq.~(\ref{eq:TimeEvolved}) to:

\begin{eqnarray}
\rho(\vec{r},t) &=&  \left\langle \psi_0 |  e^{i H t}\hat{\rho}(\vec{r}) e^{-i H t}  | \psi_0 \right\rangle = \nonumber \\
& & \frac{\int D \phi (\vec{x})  \ |\left\langle \lbrace \phi (\vec{x}) \rbrace |e^{-i H t}| \psi_0 \right\rangle |^2 |\phi (\vec{r})|^2}{\int D\phi (\vec{x}) \ | \left\langle \lbrace \phi (\vec{x}) \rbrace| e^{-i H t} | \psi_0 \right\rangle |^2}.\nonumber \\
\label{eq:appendix_avg_1}
\end{eqnarray}

In this work it will be advantageous to rewrite Eqs.~(\ref{eq:appendix_matrix_element}), (\ref{eq:appendix_action}), and (\ref{eq:appendix_avg_1}) in terms of two new fields; the density field, $\rho(\vec{x})$, and phase field $\theta(\vec{x})$. These two fields are related to $\phi(\vec{x})$ by:

\begin{equation}
\phi(\vec{x}) = \sqrt{\rho(\vec{x})}e^{i \theta(\vec{x})}
\end{equation}

\noindent In terms of the density and phase field, Eq.~(\ref{eq:appendix_avg_1}) can be written as:

\begin{eqnarray}
\rho(\vec{r},t)& &= \nonumber \\
& & \frac{\int D \rho (\vec{x}) \int D \theta(\vec{x})  \ |\left\langle \lbrace \rho(\vec{x}) \rbrace, \lbrace \theta(\vec{x}) \rbrace |e^{-i H t}| \psi_0 \right\rangle |^2 \rho(\vec{r})}{\int D \rho (\vec{x}) \int D \theta(\vec{x}) \ | \left\langle \lbrace \lbrace \rho(\vec{x}) \rbrace, \lbrace \theta(\vec{x}) \rbrace | e^{-i H t} | \psi_0 \right\rangle |^2}.\nonumber \\
\label{eq:appendix_avg_2}
\end{eqnarray}

\noindent where $| \lbrace \phi(\vec{x}) \rbrace \rangle = | \lbrace \rho(\vec{x}) \rbrace, \lbrace \theta(\vec{x}) \rbrace \rangle$. The matrix element $\left\langle \lbrace \rho(\vec{x}) \rbrace, \lbrace \theta(\vec{x}) \rbrace |e^{-i H t}| \psi_0 \right\rangle$ can also be expressed in terms of the density and phase fields:

\begin{equation}
\left\langle \lbrace \rho(\vec{x}) \rbrace, \lbrace \theta(\vec{x}) \rbrace |e^{-i H t}| \psi_0 \right\rangle =\int' D\rho(\vec{x},t) D \theta(\vec{x},t) e^{i S},
\label{eq:appendix_matrix_element_2}
\end{equation}

\noindent where the action is given by:

\begin{eqnarray}
S &=&-\int_0^t dt'\int d^2x \left[ \rho(\vec{x},t) \partial_t \theta(\vec{x},t) \right. \nonumber \\ 
&+& \frac{1}{2}\nabla \sqrt{\rho(\vec{x},t)} \cdot \nabla \sqrt{\rho(\vec{x},t)} + \frac{\rho(\vec{x},t)}{2}\nabla \theta(\vec{x},t) \cdot \nabla \theta(\vec{x},t) \nonumber \\ 
&+& \left. \frac{g}{2} \rho^2(\vec{x},t) \right]
\label{eq:appendix_action_2}
\end{eqnarray}

\noindent These manipulations are exact and do not require any approximations.

\section{Scale Invariance at the Semi-Classical Level}
\label{sec:Appendix_B}

In this appendix we examine the semi-classical solution to the dynamics and the role of scale invariance. The semi-classical solution is obtained by minimizing the action in Eq.~(\ref{eq:appendix_action_2}) and only considering the semi-classical contribution to the dynamics. This approach gives the standard hydrodynamic description of a Bose gas:

\begin{eqnarray}
0 &=& \partial_t \theta(\vec{x},t)-\frac{1}{2}\frac{\nabla^2 \sqrt{\rho(\vec{x},t)}}{\sqrt{\rho(\vec{x},t)}}+\frac{1}{2}\left(\nabla\theta (\vec{x},t) \right)^2 + g\rho(\vec{x},t), \nonumber \\
0 &=& \partial_t \rho(\vec{x},t) +\nabla \cdot (\rho(\vec{x},t) \nabla \theta(\vec{x},t)). \nonumber \\
\label{eq:appendix_hydro_eom}
\end{eqnarray}

Both of the hydrodynamic equations are independent of scale and are invariant under Eq.~(\ref{scaling}). This implies that a generic solution to this equation of motion also satisfies:

\begin{eqnarray}
\rho(\vec{x}',t',\lbrace \lambda' \rbrace) = \frac{1}{b^2}\rho(\vec{x},t,\lbrace \lambda \rbrace). \nonumber \\
\theta(\vec{x}',t',\lbrace \lambda' \rbrace) = \theta(\vec{x},t,\lbrace \lambda \rbrace).
\label{eq:appendix_scaling}
\end{eqnarray}

\noindent The set of parameters $\lbrace \lambda \rbrace$ represent any additional scales introduced by the initial conditions. These length scales explicitly break the scale invariance, and need to be rescaled alongside the spatial and temporal coordinates in the problem: $\lbrace \lambda' \rbrace = b \lbrace \lambda \rbrace$. That is, the scale invariance relates the dynamics from different initial conditions to one another.

For the remainder of this discussion we consider the case when the initial conditions introduces a single length scale into the problem, $\lambda_0$. At $t=0$, Eq.~(\ref{eq:appendix_scaling}) implies that the density field can be written as:

\begin{equation}
\rho(\vec{x},0,\lambda_0) = \frac{N}{\lambda_0^2}  f \left( \frac{x}{\lambda_0} \right),
\end{equation}

\noindent The scale invariance does not predict the function $f(x)$ but is determined by the initial conditions of the system.

\section{Coarse Grained Dynamics}
\label{sec:Appendix_C}

To understand the quantum dynamics contained in Eq.~(\ref{eq:appendix_avg_1}), we perform a coarse graining procedure. This is accomplished by splitting the fields in Eqs.~(\ref{eq:appendix_avg_1}) and (\ref{eq:appendix_action}) into long wavelength isotropic degrees of freedom, $\rho_{\lambda}(\vec{x})$ and $\theta_{\lambda}(\vec{x})$, and short wavelength anisotropic fluctuations, $\delta \rho(\vec{x})$ and $\delta \theta(\vec{x})$. This expansion is controllable in the limit of dense condensates, $\rho_{\lambda}(\vec{x}) \gg 1$, and leads to an effective theory describing the isotropic long wavelength dynamics of the system.

Motivated by the discussion in Appendix \ref{sec:Appendix_B}, we choose to work with a single parameter ansatsz for the isotropic long wavelength degrees of freedom:

\begin{eqnarray}
\rho_{\lambda}(\vec{x},t)  &=& \frac{N}{ \lambda^2(t)} f\left( \frac{x}{\lambda(t)} \right), \nonumber \\
\theta_{\lambda}(\vec{x},t) &=& \frac{x^2}{2} \frac{\dot{\lambda}(t)}{\lambda(t)} + \eta(t),
\label{eq:ansatz_appendix}
\end{eqnarray}

\noindent where $f(x)$ is a normalizable isotropic function that is regular at the origin, and $\eta(t)$ is a time dependent phase that is irrelevant to the following discussion. The parameter $\lambda(t)$ is the time dependent size of the condensate. In this approach, all the dynamical information is encoded in $\lambda(t)$, and we wish to derive an effective theory for this single parameter.

This specific form of the phase field, $\theta_{\lambda}(\vec{x},t)$, is chosen in order to satisfy the conservation law, the second line in Eq.~(\ref{eq:appendix_hydro_eom}). The dynamics of the phase field are treated semi-classically, and is of little importance for the remainder of this discussion.  However, no restrictions are placed on the density field.

Applying Eq.~(\ref{eq:ansatz_appendix}) to Eq.~(\ref{eq:appendix_action_2}) results in the following zeroth order action:

\begin{equation}
S_{\lambda}=\int_0^t dt'  \ \frac{1}{2} m \dot{\lambda}^2 +\frac{V}{2 \lambda^2}
\label{eq:appendix_action_lambda}
\end{equation}

\noindent where $m = C_1 N$ and $V = C_3 g N^2 - C_2 N$. The coefficients  $C_1$, $C_2$, and $C_3$ can be calculated once the function $f(x)$ has been specified. Some examples of these coefficients can be found in Ref.~\cite{Maki14}. 

The main effect of the anisotropic short wavelength fluctuations to the dynamics is to modify the matrix element, Eq.~(\ref{eq:appendix_matrix_element_2}), or equivalently the action, Eq.~(\ref{eq:appendix_action_lambda}). To generate the correction to Eq.~(\ref{eq:appendix_action_lambda}) we expand Eq,~(\ref{eq:appendix_action_2}) to second order in $\delta \rho(\vec{x},t)$ and $\delta \theta(\vec{x},t)$. Since the phase field is chosen to satisfy the semi-classical equation of motion, Eq.~(\ref{eq:appendix_hydro_eom}), the fluctuations $\delta \theta(\vec{x},t)$ will appear at $O\left(\delta \theta^2(\vec{x},t)\right)$, while the fluctuations in the density, $\delta \rho(\vec{x},t)$ will appear at $O\left(\delta \rho(\vec{x},t)\right)$. 

As previously noted, the separation of slow degrees of freedom and fast fluctuations also separates the isotropic and anisotropic motions. It is thus convenient to expand the fluctuations in terms of free particle eigenstates with definite angular momentum:

\begin{eqnarray}
\delta \rho(\vec{x},t) = \sum_{k,\ell \neq 0} N_{k,\ell} J_{\ell}(k x) \frac{e^{i \ell \phi}}{\sqrt{2 \pi}} \delta \rho_{k,\ell}(t) \nonumber \\
\delta \theta(\vec{x},t) = \sum_{k,\ell \neq 0} N_{k,\ell} J_{\ell}(k x) \frac{e^{i \ell \phi}}{\sqrt{2 \pi}} \delta \theta_{k,\ell}(t)
\end{eqnarray}

\noindent where $k$ and $\ell$ specify the radial mode and angular momentum, respectively, $N_{k,\ell}$ is the normalization factor associated with each radial mode, and $J_\ell(k x)$ is the Bessel function of order $\ell$. This expansion is useful as it implies that the linear coupling between the density fluctuations will in fact vanish due to the differing symmetries.

The remaining quadratic fluctuations can then be integrated out in order to derive an action in terms of the slow degrees of freedom, $\lambda(t)$. In principle there is no limitation to integrating out these fluctuations, however in practice this can be quite a challenge. In order to obtain an estimate of these fluctuations, we assume that the isotropic modes $\rho_{\lambda}(\vec{x},t)$ and $\theta_{\lambda}(\vec{x},t)$ are approximately constant over the length and time scales associated with the fluctuations. By neglecting the spatial and temporal dependence of the slow degrees of freedom, the action for the fluctuations will be diagonal in both the plane wave basis, or in the basis of definite angular momentum. An explicit calculation of the fluctuations in the plane wave basis is given in Ref.~\cite{Maki14}. 

Regardless of the basis, the effect of the fluctuations is to act as a background field upon which the long wave dynamics occur. These fluctuations  introduce a correction to the action, $\delta S$. $\delta S$ contains both real and imaginary terms. The real part of $\delta S$ renormalizes the coefficients $C_1$, $C_2$, $C_3$, and the coupling constant $g$,  while the imaginary part implies that the system under consideration has a finite lifetime. These corrections to the action are suppressed in the limit $g \ll 1$ which is the focus of this work. These corrections are thoroughly discussed in Ref.~ \cite{Maki14}.

We can now write down the final expression for the matrix element:

\begin{equation}
\langle \phi_{\lambda} | e^{-i H t} |\psi_0 \rangle = \int_{\lambda(0) = \lambda_0}^{\lambda(t) = \lambda} D \lambda(t) e^{i \int_0^t dt' \frac{1}{2} m \dot{\lambda}^2(t) + \frac{V}{2\lambda(t)^2} + i \delta S}
\label{eq:appendix_lambda_matrix_element}
\end{equation}

\noindent where $\lambda_0$ represents the size of the condensate which is in the state $|\psi_0 \rangle$ . 

Eq.~(\ref{eq:appendix_lambda_matrix_element}) is equivalent to the unitary time evolution of a wave function $\psi(\lambda, t) \equiv \langle \lambda | e^{-i H_{\lambda} t} | \lambda_0 \rangle $ under the the Hamiltonian $H_{\lambda}$:

\begin{equation}
H_{\lambda} = \frac{P_{\lambda}^2}{2m} - \frac{V}{2 \lambda^2} + i \ \text{Im} \ \delta H_{\lambda}
\label{eq:appendix_H}
\end{equation}

\noindent where $\lambda$ is now an operator with eigenstates $| \lambda \rangle$ and $P_{\lambda}$ the conjugate momentum: $\left[ \lambda, P_{\lambda} \right] = i$. $m$ and $V$ are now defined via the renormalized constants $C_1$, $C_2$, and $C_3$. The correction, $i \ \text{Im} \ \delta H_{\lambda}$, is the imaginary contribution due to the Hamiltonian from the anisotropic fluctuations:

\begin{equation}
\text{Im} \ \delta H _{\lambda} = \frac{C_4 g^2 N^2}{2 \lambda^2}.
\end{equation}

Finally, this equivalence between the full dynamics and the effective quantum mechanical model, Eq.~(\ref{eq:appendix_H}) allows one to recast Eq.~(\ref{eq:appendix_avg_2}) into the desired result:

\begin{equation}
\rho(\vec{r},t) = \frac{\int d \lambda \ \rho(\vec{r},\lambda) |\psi(\lambda,t)|^2}{\int d \lambda \ |\psi(\lambda,t)|^2}.
\end{equation}


\begin{thebibliography}{References}




\bibitem{Kadanoff66} L. P. Kadanoff, Physics {\bf 2}, 263 (1966).


\bibitem{Wilson83} K. G. Wilson, Rev. Mod. Phys. {\bf 55}, 583 (1983).


\bibitem{Sachdev} S. Sachdev, {\it Quantum Phase Transitions}, (Cambridge University Press 2011).



\bibitem{Hohenburg_Rev} P. C. Hohenberg, B. I. Halperin, Rev. Mod. Phys. {\bf 49}, 435 (1977).

\bibitem{Brown00} J. H. Brown, G. B. West (eds), {\it Scaling in Biology} (Oxford University Press, 2000).




 





\bibitem{Bosenova} E. A. Donley, N. R. Claussen, S. L. Cornish, J. L. Roberts, E. A. Cornell and C. E. Wieman, Nature {\bf 412}, 295 (2001).



\bibitem{Castin_Dum} Y. Castin and R. Dum, Phys. Rev. Lett.  {\bf 77}, 5315 (1996). 

\bibitem{Kagan96} Y. Kagan, E. L. Surkov, and G. V. Shlyapnikov, Phys. Rev. A {\bf 54}, R1753 (1996).

\bibitem{Stoof97} C. A. Sackett, H.T.C Stoof, and R. G. Hulet, Phys. Rev. Lett. {\bf 80}, 2031 (1998).











\bibitem{Mewes96} M.-O. Mewes, M. R. Andrews, N. J. van Druten, D. M. Kurn, D. S. Durfee, C. G. Townsend and W. Ketterle, Phys. Rev. Lett. {\bf 77}, 988 (1996).





\bibitem{Stringari96} S. Stringari, Phys. Rev. Lett. {\bf 77}, 2360 (1996).





\bibitem{Sengstock} S. Burger, K. Bongs, S. Dettmer, W. Ertmer, K. Sengstock, A. Sanpera, G. V. Shlyapnikov, and M. Lewenstein, Phys. Rev. Lett. {\bf 83}, 5198 (1999).


\bibitem{Hulet} K. E. Strecker, G. B. Partridge, A. G. Truscott and R. G. Hulet, Nature {\bf 417}, 150 (2002).

\bibitem{Reatto02}  L. Salasnich, A. Parola, and L. Reatto, Phys. Rev. A {\bf 66}, 043603 (2002).








\bibitem{Makotyn} P. Makotyn, C. E. Klauss, D. L. Goldberger, E. A. Cornell, and D. S. Jin, Nat. Phys. {\bf 10}, 116 (2014).

\bibitem{Radzihovsky16} X. Yin and L. Radzihovsky, Phys. Rev. A {\bf 88}, 063611 (2013); B Kain, and H. Y. Ling, Phys. Rev. A {\bf 90}, 063626 (2014); A. G. Sykes, J. P. Corson, J. P. D'Incao, A. P. Koller, C. H. Greene, A. M. Rey, K. R. A. Hazzard, and J. L. Bohn, Phys. Rev. A {\bf 89}, 021601(R) (2014); A. Rancon and K. Levin, Phys. Rev. A {\bf 90}, 021602 (2014).



\bibitem{Spivak04} R. A. Barankov, L. S. Levitov, and B. Z. Spivak, Phys. Rev. Lett. {\bf 93}, 160401 (2004).


\bibitem{Gurarie} M. S. Foster, V. Gurarie, M. Dzero, and E. A. Yuzbashyan, Phys. Rev. Lett. {\bf 113}, 076403 (2014).




\bibitem{csi_note} At the classical level, the interaction strength $g$ is merely a constant, and the system is scale invariant. However, the renormalization effect breaks the scale invariance. We work in the limit where the renormalization effects are weak. This breaking of scale invariance is discussed later.



\bibitem{Dalibard07} T. Yefsah, R. Desbuquois, L. Chomaz, K. J. Gunter, and J. Dalibard, Phys, Rev. Lett {\bf 107}, 130401 (2011).





\bibitem{Chin11} C.-L. Hung, X. Zhang, N. Gemelke, and C. Chin, Nature {\bf 470}, 236 (2011).



\bibitem{Negele_Orland} J. W. Negele and H. Orland, {\it Quantum Many-Particle Systems}, (Westview Press, 1988).


\bibitem{Feynman} R. P. Feynman, {\it Statistical Mechanics: A Set of Lectures} (W. A. Benjamin Inc, 1972).







\bibitem{Deng} S. Deng, Z.Y. Shi, Pengpeng Diao, Q. Yu, H. Zhai, R. Qi, H. Wu, Arxiv:1512.02044





\bibitem{Maki14} J. Maki, M. Mohammadi, and F. Zhou, Phys. Rev. A {\bf 90}, 063609 (2014).

\bibitem{most_probable_note} The most probable value of $\lambda$ in $|\psi(\lambda,t)|^2$ depends where the probability is concentrated. When the scattering states dominate, the most probable value is $\lambda_{sc}^{(1)}(t)$. when $gN$ is appreciable, the most probable size is $\lambda_0$, the size of the bound states. In both cases the dominant contributions to the density will come from $\lambda$ much smaller than the most probable value.




\bibitem{sc_vs_quantum_note} These anomalous contributions originate from the short distance, $k \lambda \ll m V$, structure of the effective wave function. The attractive case is discussed in the main text. For repulsive interactions, one can show that $\langle \lambda^{-2} \rangle \sim \langle \lambda_{sc}^{-2}(t) \rangle$. This implies that the density profile can be approximated by the semiclassical solution, which was observed in Refs.~\cite{Castin_Dum, Kagan96}.




\bibitem{Shick} M. Shick, Phys. Rev. A {\bf 3}, 1067 (1971).

\bibitem{Popov} V. N. Popov, Theor. Math. Phys. {\bf 11}, 565 (1972).

\bibitem{Petrov00} D. S. Petrov, M. Holzmann, G. V. Shlyapnikov, Phys. Rev. Lett. {\bf 84}, 2551 (2000).


\bibitem{Hammer04} H. W. Hammer, and D. T. Son, Phys. Rev. Lett. {93}, 250408 (2004).


\bibitem{Zhou13} M. S. Mashayekhi, J.S. bernier, D. Borzov, J.L. Song, and F. Zhou, Phys. Rev. Lett. {\bf 110}, 145301 (2013).

\bibitem{fluct_ref} When $g \log (k_0 \lambda)$ becomes appreciable, it is necessary to replace the coupling constant with $g = 2 \pi / \left( \log(a/\lambda) \right)$, where $a$ is the size of the two-body bound state in two dimensions: $a =k_0^{-1} e^{-2\pi /g}$. This specifically breaks the scale invariance and is known as a quantum anomaly. For more details see Ref.~\cite{Maki14}.


\end{thebibliography}
\end{document}